\documentclass[prb, twocolumn, preprintnumbers ,amsmath, amssymb, superscriptaddress, aps, floatfix, longbibliography]{revtex4-2}

\usepackage[left=1in,right=1in,top=1in,bottom=1in]{geometry}
\usepackage{ amssymb }
\usepackage{ xcolor }

\usepackage{mathptmx}

\usepackage{graphicx}

\usepackage{ragged2e}
\usepackage{physics}

\usepackage{accents}
\newcommand*{\dt}[1]{%
  \accentset{\mbox{\large\bfseries .}}{#1}}

\usepackage[singlespacing]{setspace}

\usepackage{tocloft}

\usepackage{fancyhdr}

\begin{document}
\title{Pseudo Electric Field and Pumping Valley Current in Graphene Nano-bubbles}
\author{Naif Hadadi}
\affiliation{Department of Physics, King Fahd University of Petroleum and Minerals, 31261 Dhahran, Saudi Arabia}

\author{Adel Belayadi}
\affiliation{Department of Physics, University of Science And Technology Houari Boumediene, Bab-Ezzouar, Algeria}
\author{Ahmed AlRabiah}
\affiliation{Department of Physics, King Fahd University of Petroleum and Minerals, 31261 Dhahran, Saudi Arabia}
\affiliation{Department of physics, Northwestern university, USA}
\author{Ousmane Ly}
\affiliation{Université de Nouakchott, Faculté des Sciences et Techniques, Département de Physique,
Avenue du Roi Faiçal, 2373, Nouakchott, Mauritanie}
\author{Collins Ashu Akosa}
\affiliation{Department of Applied Physics, School of Advanced Science and Engineering, Waseda University, Japan}
\author{Michael Vogl}
\affiliation{Department of Physics, King Fahd University of Petroleum and Minerals, 31261 Dhahran, Saudi Arabia}
\affiliation{Quantum Computing Group, IRC of Intelligent Secure systems, King Fahd University of Petroleum and Minerals,31261 Dhahran, Saudi Arabia}
\author{Hocine Bahlouli}
\affiliation{Department of Physics, King Fahd University of Petroleum and Minerals, 31261 Dhahran, Saudi Arabia}
\author{Aurelien Manchon}
\affiliation{Aix-Marseille Universit\'e, CNRS, CINaM, Marseille, France}
\author{Adel Abbout}
\email{adel.abbout@kfupm.edu.sa}
\affiliation{Department of Physics, King Fahd University of Petroleum and Minerals, 31261 Dhahran, Saudi Arabia}
\affiliation{Quantum Computing Group, IRC of Intelligent Secure systems, King Fahd University of Petroleum and Minerals,31261 Dhahran, Saudi Arabia}

\begin{abstract}
\noindent\textbf{Abstract}: The extremely high pseudo-magnetic field emerging in strained graphene suggests that an oscillating nano-deformation will induce a very high current even without electric bias. In this paper, we demonstrate the sub-terahertz (THz) dynamics of a valley-current and the corresponding charge pumping with a periodically excited nano-bubble. We discuss the amplitude of the pseudo-electric field and investigate the dependence of the  pumped valley current on the different parameters of the system. Finally, we report the signature of extra-harmonics generation in the valley current that might lead to potential modern devices development operating in the nonlinear regime. 
\end{abstract}
\maketitle 

\section{INTRODUCTION}
\begin{figure}[b!]
\centering
  \includegraphics[width=1\linewidth]{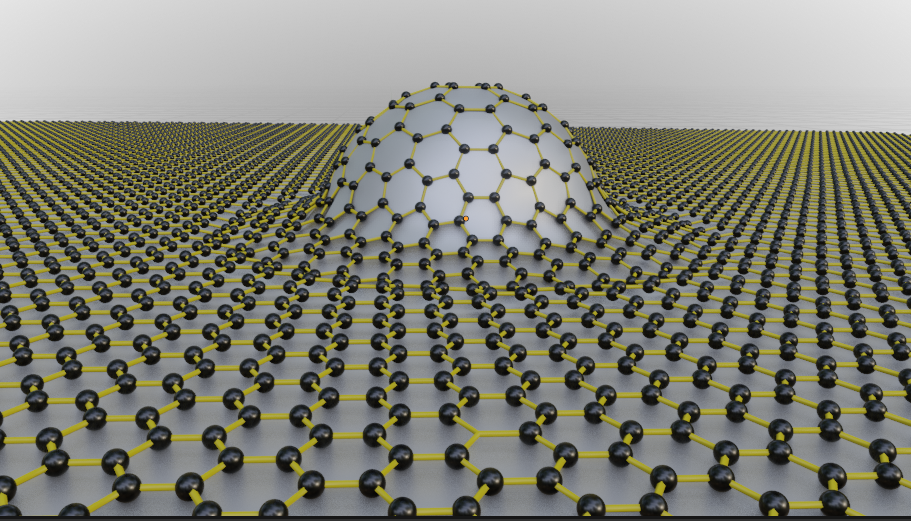}
  \caption[]{A mechanical deformation of a graphene sheet. The bump which can have different shapes (spherical, Gaussian, ...) induces an out-of-plane stretching of the hoppings between neighboring atoms. }
  \label{fig:deform}
\end{figure}

An interesting and powerful property of graphene is its ability to be stretched elastically up to 25$\%$, and to control  the induced strains in different ways \cite{masir2013pseudo,guinea2010generating}.
In addition, the unique coupling between mechanical deformation and electronic structure along with the possibility of deformation makes graphene attracts considerable attention \cite{settnes2016graphene, de2011aharonov}. This interplay was examined by observing the Landau levels that can form in graphene due to an induced strain \cite{de2011aharonov}. Indeed, deformed graphene (Fig. \ref{fig:deform})  can generate an effective gauge field $\vec{A}$ \cite{de2011aharonov}, to which one can associate a pseudo magnetic field (PMF), $
\vec{B} = \vec{\nabla} \times \vec{A}
$. This stimulated pseudo magnetic field  would allow electrons  to behave as if they were subjected to a strong real magnetic field with  strength a few orders of magnitude higher than the one generated by superconducting magnets. \cite{guinea2010energy}.


Currently, there is a great interest in generating and controlling
the valley degree of freedom of electrons in semiconductors. Indeed, the ability to manipulate valley electrons can potentially enable advanced valley-resolved electronic devices. In particular, this valley controllability opens up the possibility of using the 
momentum state of electrons, holes, or excitons as a completely new paradigm in information processing \cite{vitale2018valleytronics}. Moreover,  pure valley currents are interestingly non-dissipative currents with no accompanying net charge flow, akin to pure spin currents \cite{shimazaki2015generation}. This property is very useful in seeking ultra-low power devices.  Different routes have been proposed for generating valley currents such as using quantum pumps with a Dirac gap \cite{wang2014pure}, and using electrically induced Berry curvature in bilayer graphene \cite{shimazaki2015generation}.

In the quantum pumping technique, a cyclic change of the AC voltage of the gates  leads to the variation in the scattering matrix of the device and the generation of a DC current \cite{wang2014pure}. This method of generating a charge DC current without  bias voltage between two electrodes  \cite{brouwer1998scattering, wang2014pure} can be generalized to account for the spin or even the valley degree of freedom. In this work, a time-dependent nano-bubble in graphene, that can be created by a time-dependent voltage of an  Atomic Force Microscopy (AFM) tip, capacitively coupled to the device \cite{klimov2012electromechanical}, will be utilized as a quantum pumping device to examine the possibilities of generating a nonzero valley current with a zero net charge current. 
The deformation itself can be initially created by different techniques such as corrugated substrate engineering \cite{Bouchiat, Guinea2011} or gas inflation \cite{McEuen}.\\
The pseudo magnetic field associated with the deformation in graphene  is valley dependent and changes its sign between the two valley points K and K'.This feature allows the PMF to preserve time-reversal symmetry unlike a real external magnetic field. Consequently, a time manipulation of the PMF  will induce  different behaviour for the K and K' electrons and thus a valley-resolved study of the current is necessary \cite{Jack}. 
\textcolor{black}{It is worth mentioning that different studies have looked at the effect of other dynamical strains on the electronic transport. Ref. \cite{added1} for example, considered in-plane strain in a gapped graphene to induce a topological current, Ref. \cite{added2} looked at the symmetry requirements for pumping valley current, while Ref. \cite{added3} investigated the pseudo quantum Hall regime due to strain. Other shapes or mechanisms to get the valley filtering were also investigated \cite{, added4, AdelAbboutnew} }


\section{MODEL}
\begin{figure}[h]
\centering
  \includegraphics[scale=0.35]{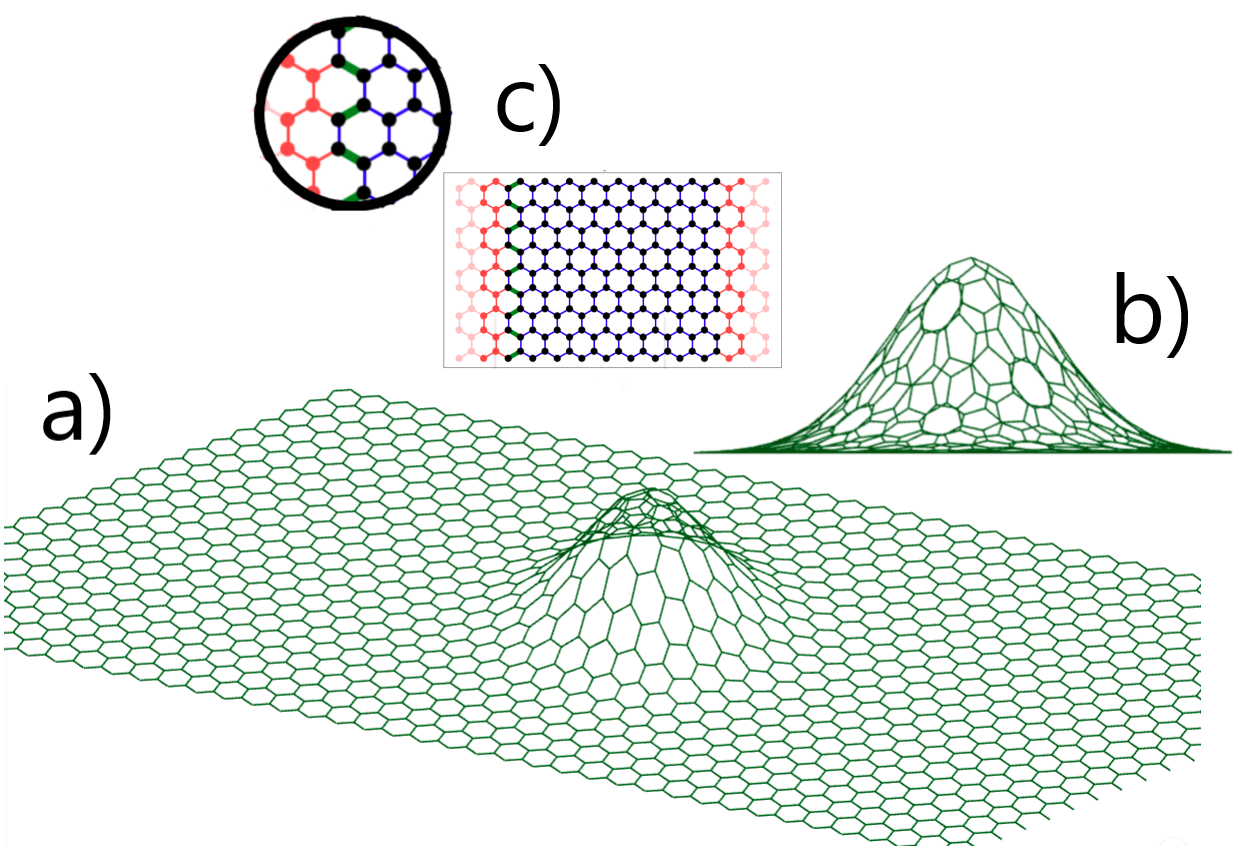}
  \caption{Tight-binding model for a quasi-1D graphene system with zigzag edges. a) shows the local deformation as a local bump in the center of the graphene sheet, altering the translational invariance of the quasi 1D waveguide. b) shows a face side of the Gaussian strain bump which is centered in the middle of the system. The green links in the zoom of  c) represent the interface through which the pumped current is calculated. The red parts in c) are the leads (reservoirs) that indicate that the system is infinite in that direction. \textcolor{black}{An animation sketching the system can be found in \cite{Supplemental} }}
  \label{fig:G_sys}
\end{figure}

 Graphene consists of carbon atoms, that are arranged in an hexagonal honeycomb structure. This arrangement is a triangular lattice with a basis of two atoms per unit cell where the lattice vector for this structure are $\va{a}_1 = \frac{a}{2}(3,\sqrt{3})$ and $\va{a}_2 = \frac{a}{2}(3, - \sqrt{3})$ (Fig. \ref{fig:G_sys}) and the constant a is  carbon-carbon distance with value $d_0$ $\sim$ 1.42 $\AA$ \cite{carrillo2014gaussian}. The electronic structure of  strained graphene is studied by utilizing a first nearest neighbor tight-binding model approximation, $\text{H} = -\sum_{\langle ij\rangle} t_{ij} c_i^\dag c_j$ where the sum runs over the nearest neighboring atoms. Figure \ref{fig:G_sys} shows the semi-infinite  graphene system, where the plane waves are coming along the zigzag direction from left and right leads. Experimentally, the deformation can be created in a graphene sheet by using an atomic force microscope's tip \cite{klimov2012electromechanical,lee2008measurement,nemes2017preparing}.  In the presence of a nano-bubble the  nearest neighbors hopping parameters become all different. Thus, the strain can be included in the system by altering the hopping parameter to the strained bond length. The adjustment of the hopping in the presence of the strain can be described by
 \begin{equation}\label{hoppingt}
t_{ij} = t_0e^{-\beta (\frac{d_{ij}}{d_0}-1)},
 \end{equation}
where $t_0 = 2.7 \text{ eV}$, $\beta = 3.37$ for graphene structure \cite{stegmann2018current}. $d_{ij}$ represents the length of the strained lattice bonds whereas $d_0$ is the lattice constant in the absence of strain.  

\section{RESULTS}
\subsection{Pseudo Magnetic Field}
\begin{figure}[t!]
\centering
  \includegraphics[width=1\linewidth]{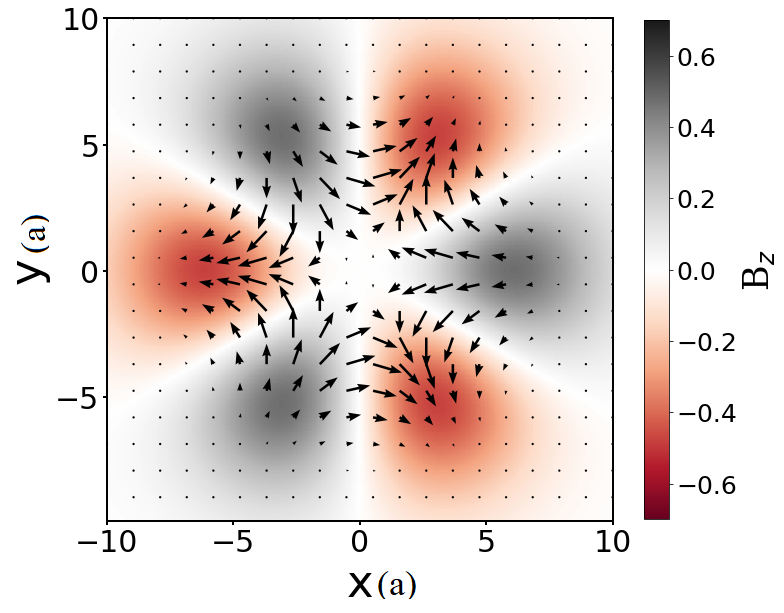}
  \caption[]{The color map indicates the threefold symmetric pseudo-magnetic field caused by the circularly symmetric Gaussian deformation. The vector field illustrates the stimulated electric field by the time-dependent pseudo-magnetic field. 'a' is the lattice constant.}
  \label{fig:EB}
\end{figure}

In our model, the graphene out-of-plane deformation is local with a characteristic width $\sigma=5 \text{ nm}$, a maximum height $h_0=3.5\text{ nm}$ \cite{de2011aharonov,parameters,settnes2016graphene} and an overall Gaussian shape modeled by $z = h_0 e^{-\frac{x^2+y^2}{2\sigma^2}}$. In the limit of low-energy carriers, the deformation is equivalent to an induced effective vector potential given by \cite{masir2013pseudo} $\vec{A} = -\frac{\hbar \beta}{2ea_0} ((\epsilon_{xx} - \epsilon_{yy})\vu*{x} - 2 \epsilon_{xy} \vu*{y})$, where the stress tensors $\epsilon_{ij}$ can be calculated by  $\epsilon_{ij} = \frac{1}{2}[(\partial_i z)(\partial_j z)]$. A straightforward calculation leads to the following form of the vector potential,


\begin{equation}
\vec{A}_0 =\frac{-\hbar \beta}{e a_{0} \sigma^{4}} z^{2}(\left(x^{2}-y^{2}\right) \hat{x}-2x y \hat{y}).
\end{equation}
The corresponding pseudo magnetic field can be obtained by $
\vec{B}_0 = \vec{\nabla} \times \vec{A}_0
$:
\begin{equation}
\vec{B}_0 = \frac{4\hbar\beta}{ea_0\sigma^6} z^2(y^3-3x^2y) \vu*{z}.
\end{equation}
\noindent The pseudo magnetic field shows alternating positive and negative regions in the deformation position as shown in Fig. \ref{fig:EB}. This magnetic field is valley dependent and changes its sign between the high symmetry points $K$ and $K^\prime$ and is responsible for the valley filtering in graphene nano-bubbles \cite{settnes2016graphene}.    

\textcolor{black}{Using eq.\ref{hoppingt}, we can demonstrate that the  maximum relative change in the hopping parameter is proportional to  $\beta h_0^2/\sigma^2$. This implies that for a small deformation width $\sigma$, it becomes necessary to  reduce the maximum height to remain within the constraints of   stretchable graphene \cite{masir2013pseudo,guinea2010generating}. In our particular case, a numerical examination of max($\Delta t_{ij}/t_{ij}$) confirms that we are, indeed, within this specified limit.}
\subsection{Induced Pseudo Electric Field}
The dynamics of the charge carriers in graphene under a dynamical external mechanical strain is a challenging problem. The generated frequencies of the perturbation projects into the sub-Terahertz domain with reported frequencies up to 200 GHz \cite{YunWu}. We note in passing that such frequencies are much smaller than the bandwidth of graphene. Therefore, standard techniques like Floquet theory \cite{Eckardt_2015,Bukov_2015,Vogl_2022,Abanin_2017,Rodriguez_Vega_2021,Vogl_2019,PhysRevB.101.024303} are hard to apply, which makes a full numerical treatment necessary and which will be the subject of our work. Things become more interesting when the strain is a nano-bubble exhibiting a few hundred Teslas pseudo magnetic field. Indeed, an oscillating bump with a height $h = h_0 + \delta h(t)$ will change the vector potential to a new form $\vec{A}=(1+\frac{\delta h(t)}{h_{0}})^2 \vec{A}_{0} $. In the Weyl gauge, the Maxwell equations show an induced pseudo electric field in the limit $\frac{\delta h(t)}{h_{0}}\ll 1$, to the lowest order:
\begin{equation}
    \vec{E}(x,y)=-2\frac{\delta \dt{h}(t)}{h_0}  \vec{A}_0,
\end{equation}
where the dot represents the time derivative. This result accounts for the small perturbations due to a small amplitude in the out-of-plane vibration of the nano-bubble which allows us to keep our interpretation within the linear response theory. It is worth noting that although the pseudo magnetic field is strong (hundreds of Teslas), the induced electric field is only a few kV/cm. This is mainly explained by the fact that these large values of $B_0$ are rather due to the large gradient of $\vec{A}_0$ (variation on very short scales) than to its magnitude. The pseudo-electric field map is depicted in Fig. \ref{fig:EB} where a threefold symmetry emerges and the system alternates between an inward and outward pseudo electric field for the $K$ and $K^\prime$ valleys.
\begin{figure}[h!]
\centering
  \includegraphics[width=0.95\linewidth]{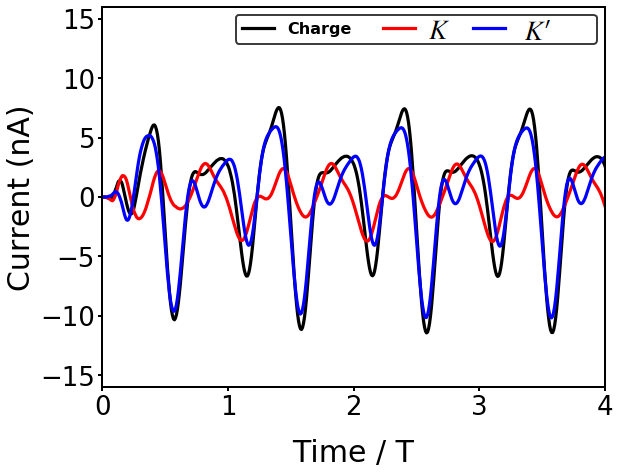}
  \caption[]{The generated currents through the interface shown in Fig. \ref{fig:G_sys}. The two types of currents at $K$ and $K^\prime$ are in phase and average to zero. The charge current is obtained for $h_0=3.5 \text{ nm}$ and $\delta h=0.35 \text{ nm}$. $E_F=0.28t_0$. T is the period of oscillation of the bubble.}
  \label{fig:currents}
  \end{figure}
  \vspace{-1em}
\subsection{Charge and Valley Currents}
As, in the case of charge-pumping by the dynamics of a skyrmion deformation \cite{AbboutSkyrmion,brouwer1998scattering} and spin-pumping by the precession of magnetic textures \cite{Brataas,Tatara}, an oscillating nano-bubble is expected to alter the charge density and induce a charge/valley current through the interface between the system and the lead \cite{Adelcomment}. To investigate this property, we adopt a nano-bubble oscillating at hundreds GHz with $\delta h(t)= \delta h_0 \sin(\omega_0 t+\phi)$ with a height modulation strength $h_0=3.5 \text{ nm}$, a characteristic width $ \sigma=5 \text{ nm}$ and a perturbation magnitude $ \delta h_0=0.35 \text{ nm}$. The way to generate these oscillations is unimportant at this stage but can be initiated by a capacitively coupled atomic force microscope's tip or a laser on top of the strained part. The first step towards the calculation of the charge current is to solve the time-dependent Schrodinger equation to obtain the wave functions at different times and then calculate the different currents in the system in one-body formalism \cite{Kloss_2021, Groth_2014}. It is important to mention that a more accurate procedure involves an integration over the different contributions from all energies in the Fermi sea. This was safely avoided due to the small driving frequencies (compared to the Fermi level $E_F=270 \text{ meV}$) and the small excitations considered in this study ($\hbar\omega/E_F\ll1$). This reduces the needed numerical resources and keeps the problem at the Fermi surface (similar to the case of the spin-pumping theory). Figure \ref{fig:currents} shows the result of the pumped charge current \textcolor{black}{$I_{\text{charge}}=I_K+I_{K'}$}, which expectedly averages to zero as explained by the scattering approach to parametric pumping. Indeed, varying one parameter is not enough to generate a sustainable charge current \cite{brouwer1998scattering}.

\begin{figure}[t!]
  \centering
  \includegraphics[height=3 cm]{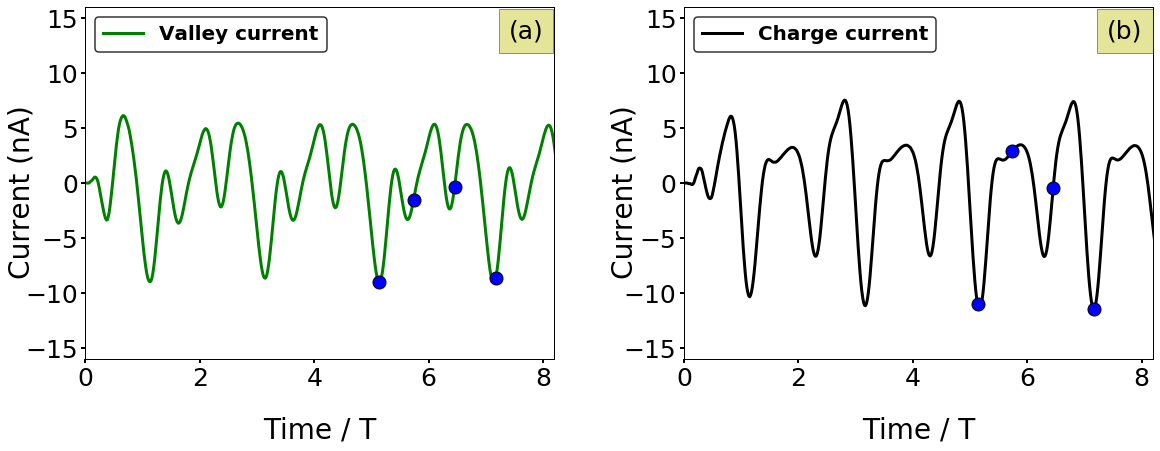}
  \caption{The charge current (Left) and the valley current (Right) pumped through an interface with the lead (see figure \ref{fig:G_sys}). The average of each of them is zero. The charge oscillates at the same frequency as the nano-bubble, whereas the pumped valley current shows higher order harmonics.\textcolor{black}{The blue dots represent the times at which the current maps were plotted in Fig. 7}. T is the period of oscillation of the nanobubble}
  \label{fig:points}
\end{figure}

\begin{figure*}[h]
  \includegraphics[width=\textwidth]{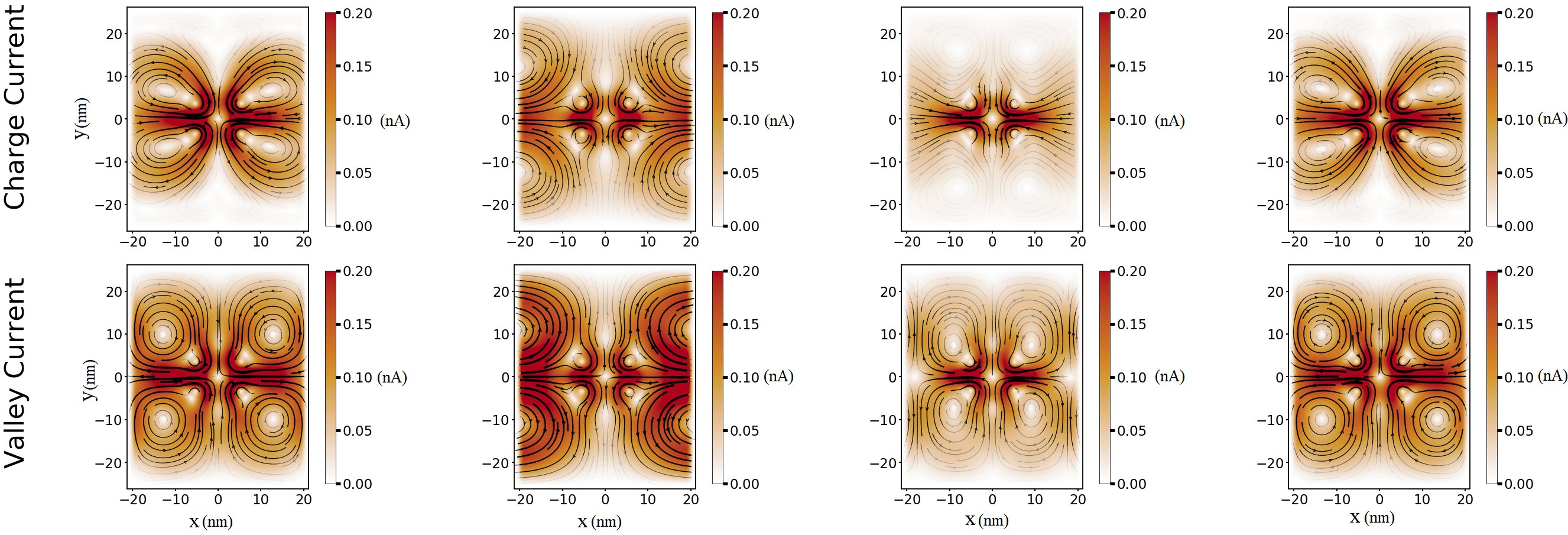}
  \caption{ The current map of the pumped charge current (top) and the valley pumped current at different times for 6 modes  with $E_{F} = 0.07t_0$. The nano-bubble size is $\sigma=5 \text{ nm}$ and the height $\delta h_0 = 0.1h_0$ . The position is expressed in $\text {nm}$.The times at which the maps are plotted are inside one period of oscillation and increase from left to right.  (the  4 different phases of an oscillation). \textcolor{black}{The blue dots in Fig. \ref{fig:points} show the time where the maps are plotted. ( for these figures, the time increases from left to right.) }  }
 \label{fig:map}
 \end{figure*}
The frequency of the signal is the same as that of the oscillating nano-bubble. So far, these results look like the same as for the spin pumping of $I_x$, $I_y$ and $I_{\text{charge}}$ in Ferromagnets (precessing around the z-axis): zero average and the same frequency as the driving perturbation. Since the pseudo electric field is valley dependent, the contribution coming from each \textit{K}-point to the charge current is investigated and plotted in Fig. \ref{fig:currents}. The two currents, $I_K$, and $I_{K^\prime}$ have amplitudes \textcolor{black}{of the same order of magnitude} and both oscillate at the same angular frequency $\omega_0$. \textcolor{black}{The two currents are not exactly the same, because for a Fermi energy slightly higher than zero, the modes in the two different valleys are non-symmetric and have slightly different wavenumbers}. The valley current, which is defined as $I_\text{valley}=I_K-I_{K^\prime}$ is of huge interest in low energy consumption devices. The reason behind it is that the net charge transferred can be zero whereas at the same time the valley current is not vanishing. This eliminates the dissipation of energy as heat while transporting information through the valleys' degrees of freedom.
The valley current is well defined in the interface region (away from the bubble), because the lead is uniform and the \textcolor{black}{conducting} modes \textcolor{black}{(incoming wave functions from the lead)} do not mix. One has to compute the time-dependent wave function for each mode, using T-kwant package \cite{Kloss_2021,Groth_2014}, and identify the different modes in each valley. The contribution of each mode to the current between sites $a$ and $b$ reads:
\begin{equation}
I^{a b}=i\left(\boldsymbol{\psi}_b^{\dagger}\left(H_{a b}\right)^{\dagger}  \boldsymbol{\psi}_a-\boldsymbol{\psi}_a^{\dagger}  H_{a b} \boldsymbol{\psi}_b\right)
\end{equation}
where, $\boldsymbol{\psi}_{a (b)}$ is the wave function of a given mode at site $ a (b)$. \textcolor{black}{$H_{a b}$ is the matrix element between sites a and b (in our case, it represents the hopping parameter between a and b in graphene)}. The contribution of modes of a given valley are summed together to finally express $I_\text{valley}$.
Inside the scattering region, this procedure can still be used, but the current is not exactly a pure valley current. Indeed, the scattering between the valleys is possible, yet it remains small (the points K and K' are far apart in the Brillouin zone) \cite{Jack}. For this reason, the approximation of the local current in the scattering region is still valid \cite{Jack}. Another approach is to project the wave functions on the valley-resolved ones of pristine graphene \cite{stegmann2018current}.

\begin{figure}%
    \centering
    \includegraphics[width=3.425cm]{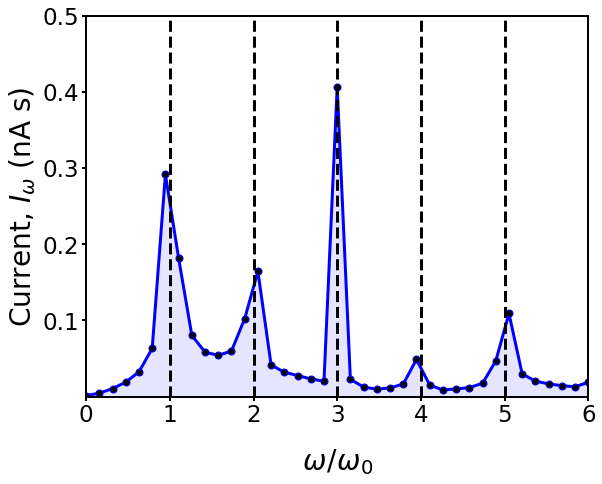}
    \qquad
    \includegraphics[width=3.23cm]{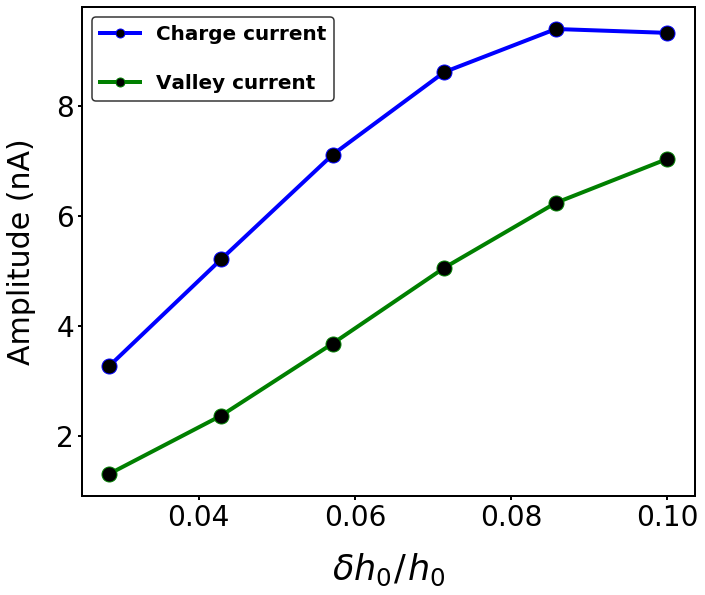} 
    \caption{ Left) The Fourier transform of the Valley current.The signal exhibits higher order frequencies (multiple of $\omega_0$). Right) (blue curve) The amplitude of the pumped electric current vs the amplitude of the bump oscillation. (green curve) The amplitude of the valley current vs the amplitude of the bump oscillation. $h_0=3.5 \text{ nm}$, $\sigma=5\text{ nm}$ and the Fermi energy is $E_F= 270 \text{ meV}$. The width of the system is $W=50 \text{ nm}$. }%
    \label{fig:example}%
\end{figure}

Fig. \ref{fig:points} shows the result for the valley current pumped through the same interface. The average is again zero and the amplitude is much smaller than that of the charge current. The most important thing to notice is that the frequency of the $I_{\text{valley}}$ is not exactly the same as that of the nano-bubble. As we can see in Fig. \ref{fig:example}, the valley current exhibits a main frequency of $3\omega_0$ with small contributions from $\omega_0$, $2\omega_0$ and $5\omega_0$. The appearance of higher-order harmonics was recently reported for spin-pumping in the presence of spin-orbit interaction \cite{Oly} or in the presence of non-collinear magnetic structure \cite{Oly2}. In our case, none of the discussed causes in \cite{Oly,Oly2} are present in our system.
\noindent The charge and valley pumping depend on the parameters of the system. As we can see it from the expression of the pseudo electric field, the latter is proportional to the perturbation $\frac{\delta h_o}{h_0}$. Naturally, this is expected to be the same for the charge and valley currents. Fig. \ref{fig:example} b) shows the amplitude of the $I_{\text{charge}}$ and $I_{\text{valley}}$ which are clearly linearly varying with the change $\frac{\delta h_o}{h_0}$. For stronger perturbations, higher non-linear effects are expected to manifest. Another way to increase the different pumped currents is to play with the frequency and use a Terahertz spectrum. This is a challenging task, but very promising results suggest that this projection is reachable \cite{YunWu}. 
\noindent The zero-average of the valley current makes it very difficult to exploit in devices since no accumulation can be obtained and only oscillating current redistribution of the valleys is achieved. With proper engineering, we still can think of it as the basis for a potentially new Field-effect transistor \cite{AdelNano}.
\subsection{Local currents}
Since the pseudo electric field is local and mainly appears in the central system, investigating the current at different regions than the interface with the lead might be of a great interest. The plot of the current map will show how the regions of the maximum pseudo magnetic field influence the current direction and generation. The wave functions are calculated at all sites of the systems which allows us to express the local currents. Fig. \ref{fig:map} illustrates the map obtained at different times during one period. The current lines for the charge and the valley are symmetric which is just the reflection of the symmetry of the Gaussian bump considered in this work. This will be different if a triaxial strain was adopted. In fact, it is known that a Gaussian deformation filters the valleys whereas the triaxial one splits them \cite{settnes2016graphene}.\textcolor{black}{ The Fermi energy is chosen low to reduce the number of modes due to the limited computing resources in this kind of 2D maps (current is calculated between each two links, at each time) and this doesn't change the general conclusions}. The maps for the valley current and the charge current are intrinsically different: for the charge current, it is hard to see the positions of the valley-dependent maxima of the pseudo-magnetic field. The situation is different for the valley current. We clearly notice loops of different orientations showing the dependence of the pseudo-magnetic field on the valley of the carriers. The size of the current's loops can be changed via the parameter $\sigma$ ie, creating very peaked  bumps or smoothly varying ones over a larger region. The number of modes for the chosen Fermi energy is very small. In order to increase the current, one needs to increase it either by changing the width of the system or by increasing the Fermi energy. \\
 \textcolor{black}{It’s worth mentioning that we did investigate deformations with elliptical symmetry,
deliberately tilting them to disrupt left-right symmetry. Unfortunately, this approach did not lead to
a sustained average valley current.}

\section*{CONCLUSION}
The pseudo magnetic field generated by a strained graphene sheet was analyzed. In the presence of a time-dependent oscillating strain, a corresponding pseudo electric field is stimulated giving rise to a redistribution of  the quantum states in momentum space, leading to charge and valley pumping. We demonstrated that the valley current exhibits extra-harmonics and that on average, the pseudo electric field did not sustain a direct current, and rather a zero average was obtained. This work opens the door to further interesting studies like the pseudo-electric field generated by the propagation of mechanical waves in graphene or the distribution of local electric currents due to phonons and temperature effects on graphene.

\section*{ACKNOWLEDGMENT}
The authors acknowledge computing time on the supercomputer SHAHEEN at KAUST Supercomputing
Centre and the team assistance. A.A. and M.V. gratefully acknowledge the support provided by the Deanship of Research Oversight and Coordination (DROC) at King Fahd University of Petroleum and Minerals (KFUPM) for funding this work through exploratory research grant No. ER221002.
\bibliography{reference}

\begin{thebibliography}{42}
\bibitem [{Ade()}]{Adelcomment}%
  \BibitemOpen
  \bibfield  {title} {\bibinfo {title} {This current might vanish on average},\
  }\href@noop {} {\ }\BibitemShut {NoStop}%
\end{thebibliography}
\end{document}